\documentstyle[preprint,aps]{revtex}

\begin{document}

\draft

\title{Monte Carlo calculations for liquid $^4$He at
negative pressure}

\author{J. Boronat and J. Casulleras}
\address{Departament de
F\'{\i}sica i Enginyeria Nuclear, Campus Nord B4-B5, \protect\\
Universitat Polit\`ecnica de Catalunya, E-08028 Barcelona,
Spain}
\author{J. Navarro }
\address{Instituto de F\'{\i}sica Corpuscular
(Centre Mixt Consejo Superior de Investigaciones Cient\'{\i}ficas
\protect\\ Universitat
de Val\`encia), Facultat de F\'{\i}sica,
Avda. Dr. Moliner 50, E-46100 Burjassot, Spain}

\maketitle

\begin{abstract}

A Quadratic Diffusion Monte Carlo method has been used to obtain
the equation of state of liquid $^4$He including the negative
pressure region down to the spinodal point. The atomic
interaction used is a renewed version (HFD-B(HE)) of the Aziz
potential, which reproduces quite accurately the features of the
experimental equation of state. The spinodal pressure has been
calculated and the behavior of the sound velociy around the
spinodal density has been analyzed.

\end{abstract}


\narrowtext

Recent experiments \cite{nissen,xiong1,petter} on cavitation in
liquid $^4$He at low temperatures have motivated the theoretical
study \cite{maris1,xiong2,maris2,solis,guilleu,jezek} of liquid
helium properties
at low temperature and negative pressure. Some interesting questions
have thus been raised, as the determination of the tensile
strength ({\it i.e.}, the magnitude of the negative pressure at
which cavitation becomes likely), and the spinodal pressure
({\it i.e.}, the pressure at which liquid helium becomes
macroscopically unstable against density
fluctuations). The standard (or classical) theory of nucleation
predicts for liquid helium \cite{aku} a tensile strength rising
from 8 atm at 2 K to 15 atm at 0.5 K, and the experiment around
1.5 K reported in Ref. \cite{nissen} seemed to confirm this
prediction. However, Maris and Xiong estimated
\cite{maris1,xiong2} the spinodal pressure to be -9 atm at 0 K,
so that the predictions of the standard theory must be
incorrect (since the liquid cannot exists for pressures lower
than the spinodal one). They also carried out an experiment
\cite{xiong1}, whose results are in contradiction with those of
Ref. \cite{nissen}, obtaining lower values for the tensile
strength.
More recently, several experiments \cite{petter}
providing information about the cavitation problem have been performed.
Unfortunately, as they do not rely on an accurate pressure calibration,
no tensile strength values are reported.

In Refs. \cite{maris1,xiong2} the spinodal pressure
was estimated by fitting to the measured \cite{abraham} sound
velocities $c$ as a function of pressure $P$
several polynomial and Pad\'e forms, and then extrapolating
into the negative pressure region to determine the zero of $c(P)$.
{}From a different point of view,
the spinodal pressure was calculated in Ref. \cite{solis} using two
different phenomenological models that reproduce the
equation of state in the measured positive pressure region. Although
an overall agreement between the phenomenological calculations and
the empirical results was obtained, some questions arise, as for
instance to what extent the extrapolated results depend on the
form used in the fit, or on the density functional used in the
calculations. It is therefore necessary to handle with a precise
equation of state for liquid $^4$He valid in the full range of
pressure, down to the spinodal one.

Many-body techniques have achieved a high level of accuracy in the
description of liquid helium. In particular, Monte Carlo (MC) methods
\cite{pisa} give exact information, apart from statistical
uncertainties, on the ground state of bosonic systems both at zero
and finite temperature. The well known interatomic interaction for
helium
has been a key ingredient to reach an excellent agreement between the
theoretical results and the experimental data.
Recently, we have used a Quadratic Diffusion
Monte Carlo (QDMC) method to calculate the equation of state in the
positive pressure region \cite{boro}. One of the main conclusions of
Ref.
\cite{boro} is that the HFD-B(HE) potential suggested by Aziz {\it et
al.} \cite{aziz2}, hereafter referred to as Aziz II potential, improves
the results obtained with the Aziz potential \cite{aziz},
especially when the density dependence of derivative magnitudes of the
energy is considered. In this work, we have extended those calculations
to lower densities, with the hope
of studying without ambiguity the zero-temperature properties
of homogeneous liquid $^4$He in this zone.

The QDMC method solves stochastically the Schr\"odinger equation in
imaginary time assuming a short-time approximation form for the Green's
function \cite{ceperley}. The ground-state wave function is sampled
in an iterative process after a time larger enough to project out
higher energy components. Rigorously, the exact ground-state energy is
obtained when the limit $\Delta t \rightarrow 0$ is considered. In
linear DMC algorithms one has to perform calculations at several
time steps, and then  extrapolate to the exact value. The QDMC method,
which has
evidenced a quadratic dependence on $\Delta t$ \cite{boro,chin},
improves the efficiency of the diffusion algorithm  making
feasible to use larger time steps than in DMC and avoiding the necessity
of the extrapolation to $\Delta t = 0$.

In order to guide the diffusion process a Jastrow
trial wave function of the form \cite{reato}
\begin{equation}
\Psi_T({\bf R}) = \prod_{i<j} \exp \left[ -\frac{1}{2}\, \left(
\frac{b}{r_{ij}} \right)^5 - \frac{L}{2} \exp \left( -
\left(\frac{r_{ij}-\lambda}{\Lambda} \right) ^2 \right) \, \right]
\label{trial}
\end{equation}
has been used in our calculations. The values of the parameters
appearing in Eq. (\ref{trial})
($L=0.2$, $\lambda=2.0\ \sigma$, $\Lambda=0.6 \ \sigma$, $b=1.20\
\sigma$
, where $\sigma = 2.556$ \AA)
have been fixed to optimize the Variational Monte Carlo (VMC) estimation
of the energy at the equilibrium density ($\rho_0=0.365\ \sigma^{-3}$).
The Monte Carlo
simulation has been carried out with 108 particles for $\rho < 0.328\
\sigma^{-3}$ and 128 particles for $\rho \geq 0.328\ \sigma^{-3}$,
maintaining an asymptotic population of 400 walkers. The errors
associated to the use of both a finite volume simulation box and a
finite population have been analyzed and, in all cases, are
smaller than the size of the inherent statistical fluctuations.

In Table I are reported the total energies (in K) obtained with the
Aziz II potential, for different values of the density (in units of
$\sigma^{-3}$). The experimental
values of Ref. \cite{bruyn} are also displayed. The origin
of the slight differences observed between theory and experiment
was discussed in Ref. \cite{boro}.

Derived quantities of the energy such as the pressure or the sound
velocity
have been obtained through a third and fourth degree interpolation,
with unnoticeable
changes when larger degrees were introduced in the interpolation method.
The QDMC prediction of $P(\rho)$ is shown in Fig. 1 (solid line) for the
whole range of densities, in
comparison with experimental data for positive pressures \cite{bruyn}.
The agreement between the Aziz II results and the experimental
data is quite impressive.

At this point, we would like to draw the attention on
the quality of the extrapolations coming from the previously available
data, laying mainly in the positive pressure region.
In the majority of microscopic calculations on liquid helium the
energy per particle is parametrized using a polynomial form
\begin{equation}
E/N = e_0 + B \, \bigg( \frac{\rho - \rho_0}{\rho_0} \bigg) ^2
+ C \, \bigg( \frac{\rho - \rho_0}{\rho_0} \bigg) ^3 \ .
\label{poli3}
\end{equation}

\noindent
On the other hand, in calculations based on Density Functional Theory,
the form
\begin{equation}
E/N = b \, \rho + c \, \rho^{1 + \gamma}
\label{strin}
\end{equation}
proposed by Stringari \cite{string}, has proved to be very efficient
in describing properties of homogeneous and inhomogeneous
(including an additional surface term) liquid $^4$He.
We have used both forms to fit our previous QDMC results \cite{boro},
which included
only one point below the equilibrium density. Proceeding in this way
({\it i.e.}, considering only the five last densities
of Table I), we have found that although
both fits are compatible with the previous results of the energy,
only the second form, see Eq. (\ref{strin}), predicts
the new QDMC
results at  densities lower than 0.328 $\sigma^{-3}$.
This fact is reflected in Fig. 1, where the pressure derived
from these fits is plotted as a function of density. The
short-dashed line corresponds to the fit (\ref{poli3}) and the
long-dashed line to
the fit (\ref{strin}). The starting points from the left of all curves
depicted in
Fig. 1 correspond to the location of the spinodal point. The QDMC result
for the spinodal pressure is $P_c=-9.30 \pm 0.15$ atm, corresponding to
a density $\rho_c=0.264 \pm 0.002\ \sigma^{-3}$.
The fit (\ref{strin}) predicts
$(\rho_c,P_c)=(0.266,-9.08)$ whereas the fit (\ref{poli3}) gives
$(\rho_c,P_c)=(0.292,-6.66)$.
This Figure illustrates
that the extrapolation from the positive pressure region to the
negative one is quite sensitive to the form used in the fit.
One can also see the capability of the built-in density
functional given in Eq. (\ref{strin}) to reproduce quite accurately the
equation of
state of liquid $^4$He at any density, even when only positive pressure
values are used as input in the numerical fit. It is also noticeable the
stability of Eq. (\ref{strin}) when all the QDMC energy results (Table
I) are
used to fix the optimal parameters $b$, $c$ and $\gamma$, being their
relative
changes less than 5 \%. In this case, the spinodal point turns out
to be $(\rho_c,P_c)=(0.267,-9.16)$. If the fit (\ref{poli3}) is
extended to all
the energy values, the spinodal point $(\rho_c,P_c)=(0.267,-9.24)$ comes
close to the Monte Carlo prediction.

In Figure 2 is displayed the sound velocity $c$ as a function of
pressure $P$. The points are the experimental values of Ref.
\cite{abraham}, and the solid line corresponds to the QDMC results.
The accuracy provided by the Aziz II potential is
again remarkable, giving results for the sound velocity in close
agreement with the experiment.
It can be seen that $c$ drops to zero very fast when approaching
the spinodal point. The behavior of $c$ near $P_c$ is expected to be of
the form $c \propto (P-P_c)^{\nu}$, being $\nu$ the critical exponent.
It is known \cite{solis} that $\nu=1/4$ provided that the quantity
\begin{equation}
   P_c^{\prime \prime}= \left( \frac{ \partial^2 P}{\partial \rho^2}
\right)_{\rho_c}
\label{pcrit}
\end{equation}
be different from zero.
Our QDMC estimation of $P_c^{\prime \prime}$ is $1200 \pm 100$
K$\sigma^3$. Therefore, the clear departure from zero of $P_c^{\prime
\prime}$ guarantees that $\nu=1/4$, in disagreement with the Maris'
model \cite{maris2} which, explicitly taking the value $P_c^{\prime
\prime} = 0$, predicts $\nu=1/3$. Nevertheless,
in the positive pressure region  the experimental values are very
well reproduced by the above form with $\nu$ close to 1/3.
In fact, we have found that for pressures
only slightly higher than $P_c$ and up to almost solidification
pressure, the behavior is also $c \propto (P-P_c)^{\nu}$ with the
exponent given by \cite{solis}
\begin{equation}
\nu =\nu_0 \equiv - P_c \,\kappa_0 \, u_0 \ ,
\label{nu0}
\end{equation}
where $\kappa_0$ is the isothermal compressibility
\begin{equation}
\kappa = \frac{1}{\rho} \, \left( \frac{\partial \rho}{\partial P}
\right)_T
\end{equation}
and $u_0$ is the Gr\"uneisen constant
\begin{equation}
u = \frac{\rho}{c} \, \frac{\partial c}{\partial \rho} \ ,
\end{equation}
both quantities evaluated at the saturation density $\rho_0$.
Our QDMC results give for these quantities the values
$\kappa_0 = 0.01231 \pm 0.00005$ atm$^{-1}$ and $u_0 = 2.81 \pm
0.04$ with $\rho_0=0.3645 \pm 0.0003\ \sigma^{-3}$, to be compared with
the experimental values \cite{abraham} $\kappa_0^{\rm exp}=0.01230$
atm$^{-1}$, $u_0^{\rm exp}=2.84$ and $\rho_0^{\rm exp}=0.3646\
\sigma^{-3}$. The QDMC result for the exponent (\ref{nu0}) is
$\nu_0=0.322 \pm 0.007$, close to the value $1/3$ pointed out in
Refs. \cite{xiong2,maris2}.

The power-law behavior of $c$ is shown in Figures 3 and 4. The
quantity $(c/c_0)^{1/\nu}$, where $c_0$ is the sound velocity at
the equilibirium density, is displayed in Figure 3 against $1
- P/P_c$. When $\nu=\nu_0$
a straigth line is obtained in a very wide range of pressures (Fig.
3), whereas the case $\nu=1/4$ manifests a nearly quadratic behavior.
However, for pressures close to the spinodal pressure one can see
(Fig. 4) that the linear behavior is obtained taking the critical
exponent as $1/4$.

To summarize, we have calculated the equation of state of liquid
$^4$He in a QDMC framework using the Aziz II potential from the
spinodal point up to the solidification point. The
characteristics of the spinodal point have been evaluated
without the uncertainty of the extrapolation from the positive
pressure data. As a byproduct, we have noted that our QDMC
results are very well fitted by the form (\ref{strin}), giving
thus additional support to this purely phenomenological density
functional.

This work has been partly supported by DGICyT (Spain) Grant Nos.
PB90-06131, PB92-0761 and  PB92-082. Most of the simulations were
performed on a multiprocessor CM2 of the CEPBA (Centre Europeu de
Paral$\cdot$lelisme de Barcelona).

\begin{figure}
\caption{Pressure as a function of density.
Points: experimental results \protect\cite{bruyn}, solid line:
QDMC result, short-dashed line: fit (\protect\ref{poli3}), and
long-dashed line: fit (\protect\ref{strin}).
The fits (\protect\ref{poli3}) and (\protect\ref{strin}) include only
the results from
.328 to .438 $\sigma^{-3}$.}
\end{figure}

\begin{figure}
\caption{The sound velocity as a function of
pressure. The experimental points are taken from Ref.
\protect\cite{abraham}, and the solid line corresponds to the QDMC
results.}
\end{figure}

\begin{figure}
\caption{Power-law behavior of the sound velocity using
as a critical exponent $\nu_0$ (\protect\ref{nu0}) and $1/4$.}
\end{figure}

\begin{figure}
\caption{Same as Figure 3 for pressures close to the
spinodal one. To illustrate more clearly the power-law behavior, the
quantity $(c/c_0)^{1/\nu}\, / \, (1-P/P_c)$ has been plotted against
$1-P/P_c$ for the values $\nu=\nu_0$ and $1/4$.}
\end{figure}

\begin{table}
\caption{Results for the total energies (in K) at
several densities (in $\sigma^{-3}$)
from the QDMC calculations with the Aziz II potential and
experiment \protect\cite{bruyn}.}
\begin{tabular}{ccc}
 $\rho$ &  $E/N$ & $E/N$ (exp) \\ \tableline
0.264 &  -6.376 $\pm$ 0.010 & \\
0.267 &  -6.441 $\pm$ 0.010 & \\
0.285 &  -6.692 $\pm$ 0.010 & \\
0.300 & -6.892 $\pm$ 0.010 & \\
0.310 & -7.011 $\pm$ 0.010 & \\
0.328 & -7.150 $\pm$ 0.010 & \\
0.365 & -7.267 $\pm$ 0.013 & -7.17 \\
0.401 & -7.150 $\pm$ 0.016 & -7.03 \\
0.424 &  -6.877 $\pm$ 0.022 & -6.77 \\
0.438 &  -6.660 $\pm$ 0.017 & -6.55 \\
\end{tabular}

\end{table}

\end{document}